\documentclass[reprint,amsmath,amssymb,prd]{revtex4-1}
\usepackage{dcolumn}
\usepackage{bm}
\usepackage{mathrsfs}
\usepackage{balance} 
\usepackage{graphicx}

\begin{document}

\title{Gravitational Spin Hall Effect of Dirac Particle and the Weak Equivalence Principle}
\author{Zhen-Lai Wang}
\email{wangzhenlai@hbpu.edu.cn}
\affiliation{Center for Fundamental Physics and School of Mathematics and Physics,
Hubei Polytechnic University, Huangshi 435003, China}
\date{\today}
\begin{abstract}
We present a spin-induced none-geodesic effect of Dirac wave packets in a static uniform gravitational field. Our approach is based on the Foldy-Wouthuysen transformation of Dirac equation in a curved spacetime, which predicts the gravitational spin-orbit coupling. Due to this coupling, we find that the dynamics of the free-fall Dirac wave packets with opposite spin polarization will yield the transverse splitting in the direction perpendicular to spin orientation and gravity, which is known as the gravitational spin Hall effect. Even in a static uniform gravitational field, such effect suggests that the weak equivalence principle is violated for quantum particles.
\end{abstract}

\maketitle

\section{Introduction}

The question whether the weak equivalence principle (WEP) also holds for quantum particles has been received an theoretical and experimental interest for a long time~\cite{Cole,Alva,Viol,Pete,Fray,Chow,Rosi,Orla,Schw,Flor}. The WEP is one of the foundational assumption of classical gravitational theory, which states that free-fall point particles will follow mass-independent trajectories. In classical physics the WEP is well-defined in terms of spacetime trajectories, but in quantum physics it is ill-defined because trajectories and even point particles are vague concepts. Quantum particles bring in many new properties distinct from classical point-like particles, such as matter/anti-matter~\cite{ALPH,Hohe,Hami}, spin~\cite{Ni10,Tara,Duan} and internal structures\cite{Fray,Rosi,Schl}, which might raise objections to the validity of the WEP and add new physics contents to the WEP. Furthermore, WEP-test based on quantum particle can offer vital clue to understanding the connection between the quantum and gravitational theories. For example, almost all theories trying to unify gravitational theory and the standard model of particle physics predict the violations of WEP~\cite{Damo,Lamm}.

The investigations on the WEP of quantum system have a vast spectrum, from the effects described by the simplest Schr{\"o}dinger equation with gravitational potential to the effects originated from the quantization of gravitational field~\cite{Davi,Dalv,MAli,Acci,Mous,Tino}. However, theoretical literature frequently offers conflicting views on whether the WEP of quantum system is violated or not. The major reason for this unpleasant situation is that there is no consensus about quantum version of the WEP's notion, which is still an open issue~\cite{Herd,Okon,Zych}. Quantum physic is formulated on the distinct concepts (such as quantum states, measurements and probabilities) from the classical ones. Naturally, the notions of quantum WEP can not be transferred directly from the classical statements of WEP and ought to be explained in the language of these quantum concepts~\cite{Seve17,Anas18}.

In this paper, we take the notions of quantum WEP introduced recently in Ref.~\cite{Seve17} and \cite{Anas18}. In Ref.~\cite{Seve17}, the notion of quantum WEP is reconstructed as the statement that \textit{the Fisher information about the mass of quantum probe in free fall is the same as the free case without gravitational field}. To extract information of mass through measurements of quantum probe's position, the Fisher information can be defined as 
\begin{equation}\label{fi}
F_x(m)=\int \mathrm{d}\bm x~|\psi(\bm x,t )|^2\big[\partial_m \mathrm{log}|\psi(\bm x,t)|^2\big]^2
\end{equation}
with the wave function $\psi(\bm x,t)$ of the quantum probe. The notion of quantum WEP based on Fisher information is $F_x^g(m)=F_x^f(m)$, where $F_x^g(m)$ is the Fisher information in free fall and $F_x^f(m)$ in the free case. In other words, the gravitational field can not create more information about the mass of free-fall quantum probe. 

In Ref.~\cite{Anas18}, the notion of quantum WEP is put into the following statement: \textit{The probability distribution of position for a free-falling particle is the same as the probability distribution of a free particle, modulo a mass-independent shift of its mean}. For a static uniform gravitational field, both the above the notions of quantum WEP are valid with taking no account of the internal degrees of freedom of quantum probe. However, both notions are invalid in the complex gravitational fields such as gravitational gradient field~\cite{Seve17} and gravitational wave~\cite{Quac20}.

One of the interesting and important questions is that how much the spin of quantum particle is important on these two notions of quantum WEP. For this reason, we consider the dynamics of a Dirac particle freely falling in a static uniform gravitational field. An interesting effect is revealed that the free-fall Dirac wave packets with opposite spin polarization are split transversely in the direction perpendicular to spin orientation and the gravitational acceleration. Such effect is known as the gravitational spin Hall effect~\cite{Oance19}. Like the spin Hall effects in other realms of physics, the gravitational spin Hall effect is also originated from the spin-orbit interaction~\cite{Bliok15}.

The gravitational spin Hall effects or similar effects have been reported in various physical systems such as the Dirac field~\cite{Audr81,Rudi81,Goss07,Cian08,Oanc23}, the electromagnetic/light waves~\cite{Gosse07,Duva17,Oanc20,Frol20,Hart22,Lian22} and gravitational waves~\cite{Yama18,Ande21}. Most of the studies present the polarization- or helicity-dependent ray trajectories of motion which suggest the gravitational spin Hall effect. However, many results of these studies are inconsistent with each other due to the dependence upon different methods or models. Besides, the similar effects of Dirac field discussed in the literature, the gravitational spin-orbit coupling are mainly in the charge of interaction between spin and gravity through the  gradient or torsion of the gravitational field.

In this paper, with the help of Foldy-Wouthuysen transformation~\cite{Fold}, a gravitational spin-orbit coupling is derived from the Dirac equation in a static uniform gravitational field (without gravitational gradient and torsion), which also yields the gravitational spin Hall effect even the gravitational field is so simple. Additionally, unlike the semi-classical approach where the external (position) degrees of freedom treated as a classical variable, we analyze the dynamical evolution of the wave packets of Dirac particle and show the entanglement between the internal (spin) and external (position) degrees of freedom in its full quantum-mechanical form. More importantly, we find that both the notions of quantum WEP mentioned earlier are unacceptable to the gravitational spin Hall effect presented in this paper, and so the gravitational spin Hall effect of Dirac particle can be treated as a new probe of quantum WEP's notion. 

The content of the paper is organized as follows: First, using the standard Foldy-Wouthuysen (F-W) transformation leads to a gravitational spin-orbit coupling from the Dirac equation in a static uniform gravitational field. Then, from the Schrödinger equation in the F-W picture, we consider the dynamical evolution of the Dirac wave packets, and find that the gravitational spin-orbit coupling plays a key role in the gravitational spin Hall effect going against the notions of quantum WEP. Finally, discussion and summary are presented.

\section{Dirac Equation in a Static Gravitational Field and Foldy-Wouthuysen Picture}\label{sec II}

\begin{figure}[ht]
	\centering
\includegraphics[width = 0.45\textwidth]{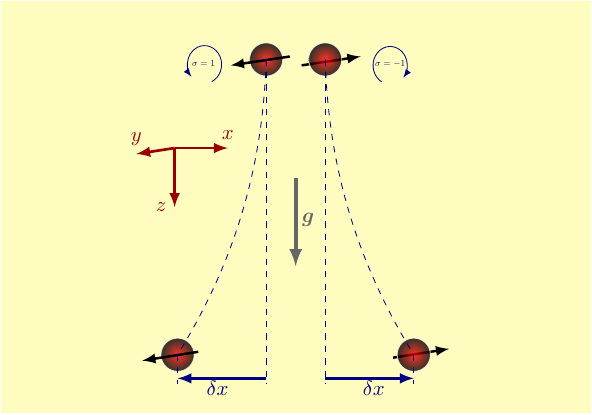}
	\caption{Schematic diagram to display gravitational spin Hall effect  of Dirac particle. Static Dirac particles carrying spin along the $y$ axis fall freely in a static uniform gravitational field. The gravitational acceleration $\bm{g}$ is along the $z$ axis. The splitting of Dirac particles with opposite spin polarization occurs in the $x$-direction.} \label{pic1}
\end{figure}
 Let us start with Dirac equation in a curved spacetime~\cite{Carr}, which describes the dynamics of a spin-1/2 particle in a gravitational field ($\hbar=c=1$),
 \begin{equation}\label{de}
	(i{\gamma}^{a}D_{a}-m)\psi=0.
\end{equation}
Hereafter Latin indices, $(a,b,c\ldots)$, denote local flat-spacetime indices and run from $0$ to $3$. Greek indices, $(\mu,\nu,\ldots)$ denote curved-spacetime indices and run from $0$ to $3$. Three-vectors are denoted by bold letters and their components are labeled by Latin indices from the middle of the alphabet, $(i,j,k,\ldots)$. Einstein summation convention is used. The spacetime metric $g_{\mu\nu}$ and the local flat metric $\eta_{ab}$ are connected by the tetrad field $g_{\mu\nu} =e_{\mu}^{~a}e_{\nu}^{~b}\eta_{ab}$, which satisfies the orthogonality conditions $e^{\mu}_{~a}e_{\nu}^{~a}=\delta^{\mu}_{~\nu}$, $e_{\mu}^{~a}e^{\mu}_{~b}=\delta^{a}_{~b}$. We adopt the flat metric $\eta_{ab}=\rm{diag}(+,-,-,-)$. The covariant derivative $D_{a}$ for the spinor field is given by
\[
D_{a}={\rm e}^{\mu}_{~a}D_{\mu},~~~~D_{\mu}=\partial_{\mu}-\frac{i}{4}\omega_{\mu}^{~ab}\sigma_{ab}.
\]
Here $\sigma_{ab}=i[\gamma_a,\gamma_b]/2$ with $[...~,...]$ denoting the commutator, and the spin connection obey $\omega_{\mu}^{~ab}=-\omega_{\mu}^{~ba}$ which can be expressed as 
\[
\omega_{\mu}^{~ab}=e_{\nu}^{~a}e^{\lambda b}\Gamma_{\mu\lambda}^{\nu}-e^{\lambda b}\partial_\mu e_{\lambda}^{~a}
\]
 with the affine connection $\Gamma_{\mu\lambda}^{\nu}$ and the tetrad field. The affine connection is constructed by the so-called Christoffel symbol $\Gamma_{\mu\nu}^{\lambda}=g^{\lambda\sigma}(\partial_\mu g_{\nu\sigma}+\partial_\nu g_{\sigma\mu}-\partial_\sigma g_{\mu\nu})/2$ from the spacetime metric $g_{\mu\nu}$. Throughout this paper, we choose the Dirac representation of gamma matrices: 
\[
\gamma^{0}=\begin{pmatrix}
	I&0\\
	0&-I
\end{pmatrix},~~~\gamma^{i}=\begin{pmatrix}
	0&\sigma^i\\
	-\sigma^i&0
\end{pmatrix},~ i=1,2,3
\]
where $\sigma^i$ is the Pauli matrices.

In this paper, we discuss only a Dirac particle traveling in a static uniform gravitational field. Following the notion of~\cite{Obuk01,Sile05}, we consider the static uniform spacetime
\begin{equation}
ds^2=V^2(dt)^2-d{\bm x}\cdot d\bm{x}
\end{equation}
where
\begin{equation}\label{metric}
	V=1-gz.
\end{equation} 
Here the gravitational acceleration $g$ along the positive direction of $z$-axis. Under such metric, Eq.(\ref{de}) can be rewritten in a Schr{\"o}dinger-like form $i\partial_t\psi=H\psi$. The Hamiltonian has the form
\begin{equation}\label{Ha1}
 H=\beta mV+\frac{1}{2}\{V,{\bm \alpha}\cdot{\bm p}\}.
 \end{equation}
Here $\alpha^{i}=\gamma^{0}\gamma^{i}$, $\beta=\gamma^{0}$, ${\bm p}=-i\hbar\nabla$ and $\{...~,...\}$ denotes the anticommutator.

Strictly speaking, the Dirac equation is self-consistent only in the context of quantum field theory due to creation and annihilation of the particle-antiparticle pairs. Fortunately, the effect of particle-antiparticle pair creation or annihilation is negligible in the non-relativistic limit of the Dirac Hamiltonian. In order to obtain the non-relativistic physical interpretation of the Dirac Hamiltonian, there is a need of carrying out the F-W transformation, which is a systematic way of separating the positive and the negative energy states. In the F-W representation, the Hamiltonian and all operators are block-diagonal and the operators are exactly analogous to the ones in the non-relativistic quantum theory.

We start from the Hamiltonian (\ref{Ha1}) in the form
\begin{equation}
 H=\beta m+ \mathcal{E}+ \mathcal{O}
 \end{equation}
with 
\begin{equation}\label{EO}
\mathcal{E}=\beta m(V-1), ~~\mathcal{O}=\frac{1}{2}\{V,{\bm \alpha}\cdot{\bm p}\},
 \end{equation}
where $\mathcal{E}$ denotes the even operator commuting with $\beta$, $\mathcal{E}\beta=\beta \mathcal{E}$, and $\mathcal{O}$ denotes the odd operator, $\mathcal{O}\beta=-\beta \mathcal{O}$. 

After two successive standard F-W transformations, the Dirac Hamiltonian can be put into the form~\cite{Bjor}
\begin{equation}
{\mathcal H}_{\scriptscriptstyle{FW}}=\beta m +\mathcal{E}+\frac{1}{2m}\beta\mathcal{O}^2-\frac{1}{8m^2}[\mathcal{O},[\mathcal{O},\mathcal{E}]]+O(g^2).
\end{equation}
Substituting Eq.(\ref{EO}) explicitly and performing the operator products to the first order of $g$ and $1/m$, we obtain
\begin{equation}\label{Hfw}
{\mathcal H}_{\scriptscriptstyle{FW}}=\beta V(m+\frac{{\bm p}^2}{2m})+\frac{i\beta g}{2m}p_3
-\frac{\beta \bm\Sigma\cdot(\bm g\times \bm p)}{4m},
\end{equation}
where $\bm\Sigma=(\sigma^{23},\sigma^{31},\sigma^{12})$ is the spin operator (or $\Sigma_k=\varepsilon_{ijk}\sigma^{ij}/2$). This F-W Hamiltonian is in agreement with~\cite{Hehl90}, and confirms the weak-field and non-relativistic approximation of~\cite{Sile05}. The last term on the right-hand side of (\ref{Hfw}) describe a gravitational (inertial) spin-orbit coupling, first introduced in~\cite{Hehl90}. This coupling plays a crucial role in determining gravitational spin Hall effect of Dirac wave packets in a static uniform gravitational field, as we will discuss in the following section.

\section{Gravitational Spin Hall Effect of Dirac Particle and the Weak Equivalence Principle}\label{sec III}
In this core section of this paper, we analyze the full quantum dynamics of the gravitational spin Hall effect of Dirac particle in terms of quantum-mechanical postulates. We start from the Schr{\"o}dinger equation in the F-W representation (\ref{Hfw})
\begin{equation}\label{SE}
i\partial_t\psi(\bm x,t)={\mathcal H}_{\scriptscriptstyle{FW}}\psi(\bm x,t).
\end{equation}
Since ${\mathcal H}_{\scriptscriptstyle{FW}}$ independent of time, the general solution to Eq.(\ref{SE}) can be represented as
\begin{equation}\label{fwave}
\psi(\bm x,t)=U(t,0)\psi(\bm x,0),
\end{equation}
as the initial wavefunction $\psi(\bm x,0)$ is known. The wavefunction at some later time $t$ is governed by the time-evolution operator $U(t,0)=U=e^{-i{\mathcal H}_{\scriptscriptstyle{FW}}t}$ on the initial wavefunction. 

Our motivation is to look for whether the WEP is violated if considering spin effect of Dirac particles in a gravitational filed. For this purpose, let us consider Dirac particles of opposite spin-polarization with the same initial mean position and momentum falling freely in our static uniform gravitational field and compare their evolution. For concreteness, the initial states of particles are represented by the following normalized gaussian wave packets, respectively,
\begin{equation}\label{wf}
\psi_{\pm}(\bm x,0)=\phi(\bm x,0)\chi_{\pm}
\end{equation}
with the normalized gaussian wave packet
\begin{equation}
\phi(\bm x,0)=(a^2\pi)^{-3/4}~\mathtt{Exp}\Big[-\frac{\bm x^2}{2 a^2}\Big]
\end{equation}
and the spinors
\begin{equation}
\chi_{+}=\frac{(1~~ i ~~0 ~~0)^T}{\sqrt{2}},~~\chi_{-}=\frac{(i~~ 1 ~~0 ~~0)^T}{\sqrt{2}}.
\end{equation}
Here the letter $T$ represents the standard transposition operation. $\psi_{+}(\bm x,0)$ and $\psi_{-}(\bm x,0)$ describe the particle of spin along the positive and negative $y$-direction initially well localized in position and momentum with the expected values $\langle\bm x \rangle=0$ and $\langle \bm p \rangle=0$, respectively, as depicted in Fig.(\ref{pic1}). 

We turn now to the  calculation of the expected position of the wave packets of opposite spin-polarization in a uniform gravitational field:
\begin{equation}\label{pos}
\langle \bm x \rangle=\big<\psi(\bm x,t)|\bm x|\psi(\bm x,t)\big>=\big<\psi(\bm x,0)|U^{\scriptscriptstyle{-1}}{\bm x} U|\psi(\bm x,0)\big>.
\end{equation}
There is a need for calculating the operator $U^{\scriptscriptstyle{-1}}{\bm x} U$, but it is difficult to solve the exact solution because of the non-commutation of operators. If considering the weak-field and non-relativistic approximation, $U^{\scriptscriptstyle{-1}}{\bm x} U$ can be written via Baker-Hausdorff lemma~\cite{Saku} as 
\begin{equation}\label{Bch}
U^{\scriptscriptstyle{-1}}{\bm x} U={\bm x}+it[{\mathcal H}_{\scriptscriptstyle{FW}},\bm x]-\frac{t^2}{2}[{\mathcal H}_{\scriptscriptstyle{FW}},[{\mathcal H}_{\scriptscriptstyle{FW}},\bm x]]+\cdots
\end{equation}
Using (\ref{Hfw}) and keeping  operator products to the first order of $g$ and $1/m$, we can deduce the following commutation rules: 
\begin{align}
&[{\mathcal H}_{\scriptscriptstyle{FW}},x]=-\frac{i\beta V p_1}{m}+\frac{ig\beta\Sigma_2}{4m},\\
&[{\mathcal H}_{\scriptscriptstyle{FW}},y]=-\frac{i\beta V p_2}{m}-\frac{ig\beta\Sigma_1}{4m},\\
&[{\mathcal H}_{\scriptscriptstyle{FW}},z]=-\frac{i\beta V p_3}{m}+\frac{g\beta}{2m},\\
&[{\mathcal H}_{\scriptscriptstyle{FW}},[{\mathcal H}_{\scriptscriptstyle{FW}},x]=[{\mathcal H}_{\scriptscriptstyle{FW}},[{\mathcal H}_{\scriptscriptstyle{FW}},y]=0,\\
&[{\mathcal H}_{\scriptscriptstyle{FW}},[{\mathcal H}_{\scriptscriptstyle{FW}},z]=-g.
\end{align}
Substituting these commutation rules back into Eq.(\ref{Bch}), and performing the integration of (\ref{pos}) with the initial wave packets (\ref{wf}), we can get 
\begin{gather}
\langle {x}(t) \rangle_{+}=-\frac{gt}{4m},~\quad~\langle {x}(t) \rangle_{-}=\frac{gt}{4m}\\
\langle {y}(t) \rangle_{+}=\langle {y}(t) \rangle_{-}=0\\
\langle {z}(t) \rangle_{+}=\langle {z}(t) \rangle_{-}=\frac{1}{2}g t^2
\end{gather}
Such results can be rewritten in a compact form of 3-vectors as 
\begin{equation}\label{SHE}
\langle {\bm x}(t) \rangle_{\pm}=(-\frac{\sigma gt}{4m},~0,~\frac{1}{2}g t^2),
\end{equation}
where $\sigma=+1$ for particle's spin along the positive y-direction and $\sigma=-1$ for particle's spin along the negative y-direction, respectively. Without considering the quantum fluctuations, the classical trajectory ${\bm x}(t) =\bm{g} t^2/2$ is recovered in the semi-classical limit $\hbar\to 0$.

From Eq.(\ref{SHE}), we can find that the wave packets of opposite spin-polarized Dirac particles in free fall can be split transversely in the direction perpendicular to spin and gravity, as illustrated in Fig.(\ref{pic1}), which is known as the gravitational spin Hall effect. Free-fall Dirac particles with different spin orientations follow `different paths'. Besides, unlike the classical case, the expected trajectory of the quantum particle in free fall is dependent on its mass, and the quantum particle with spin polarization is able to fall freely in a different ‘path structure’. In these senses, this effect implies a kind of violations of quantum WEP induced by the spin of quantum particle. 

Restoring explicit factors $\hbar$ and $c$, we can get the transverse shift from the origin point along $x$-axis by an amount 
\begin{equation}
\delta x\sim\frac{\hbar gt}{4mc^2}=\frac{\hbar}{4mc}\frac{gt}{c}\sim\frac{\lambdabar}{4}\frac{\langle v \rangle}{c}=\frac{\lambdabar}{4}\frac{\sqrt{2g{\langle z \rangle}}}{c}
\end{equation}
Here, $\lambdabar$ is the reduced Compton wavelength. Although this effect originates from the wave feature of quantum particle, the split is due to the geodesic deviation out of gravitational (inertial) spin-orbit coupling, existing even in a uniform gravitational field. In addition, this effect implies the entanglement between the internal (spin) and external (position) degrees of freedom of the wave packets.

In order to analyze the detail dynamical evolution of wavefunction and the detail entanglement of the internal and external degrees of freedom, we can try to look for the solution (\ref{fwave}) of the wavefunction at time $t$. The F-W Hamiltonian can be written as
\begin{equation}
{\mathcal H}_{\scriptscriptstyle{FW}}={\mathcal H}_{f}+{\mathcal H}_{g}	
\end{equation}
with defining the free Hamiltonian
\begin{equation}
{\mathcal H}_{f}=\beta m+\beta\frac{{\bm p}^2}{2m}	
\end{equation}
and the interaction Hamiltonian
\begin{equation}
{\mathcal H}_{g}=-\beta mgz-\frac{\beta g}{4m}\{{\bm p}^2,z\}-\frac{\beta\bm\Sigma\cdot(\bm g\times \bm p)}{4m}
\end{equation}
We can deduce the following commutation rules to the first order in term of $1/m$ and $g$:
\begin{equation}
[{\mathcal H}_{f},{\mathcal H}_{g}]=igp_{3},[{\mathcal H}_{g},[{\mathcal H}_{f},{\mathcal H}_{g}]]=[{\mathcal H}_{f},[{\mathcal H}_{f},{\mathcal H}_{g}]]=0
\end{equation}
Using these commutation rules and the Baker-Campbell-Hausdorff (BCH) formula~\cite{Serr}, we can factorize the time evolution operator as
\begin{equation}
U=e^{-i{\mathcal H}_{f}t-i{\mathcal H}_{g}t}\sim e^{-i{\mathcal H}_{g}t}e^{-\frac{1}{2}[{\mathcal H}_{f},{\mathcal H}_{g}]t^2}e^{-i{\mathcal H}_{f}t}
\end{equation}
Thus, the evolution of the wavefunction will become
\begin{align}\label{wf1}
\psi_{\pm}(\bm{x},t)&\sim e^{-i{\mathcal H}_{g}t}e^{-\frac{1}{2}[{\mathcal H}_{f},{\mathcal H}_{g}]t^2}e^{-i{\mathcal H}_{f}t}\psi_{\pm}(\bm{x},0)\nonumber\\
&=e^{-i{\mathcal H}_{g}t}e^{-\frac{1}{2}[{\mathcal H}_{f},{\mathcal H}_{g}]t^2}\psi_{f\pm}(\bm{x},t)\nonumber\\
&=e^{-i{\mathcal H}_{g}t}e^{-\frac{1}{2}gt^2\partial_3}\psi_{f\pm}(\bm{x},t)\nonumber\\
&=e^{-i{\mathcal H}_{g}t}\psi_{f\pm}(\bm{x}-\frac{1}{2}{\bm g}t^2,t)
\end{align}
Here $\psi_{f\pm}(\bm{x},t)=e^{-i{\mathcal H}_{f}t}\psi_{\pm}(\bm{x},0)=e^{-imt}e^{-it\frac{{\bm p}^2}{2m}}\psi_{\pm}(\bm{x},0)$ describes the free evolution of the initial wavefunction $\psi_{\pm}(\bm{x},0)$ without gravitational interaction, which can be written explicitly as~\cite{Grif}
\begin{equation}\label{wff}
\psi_{f\pm}(\bm{x},t)=\phi_{f}(\bm{x},t)\chi_{\pm}
\end{equation}
where 
\begin{equation}
\phi_{f}(\bm{x},t)=\Big(\frac{a^2}{\pi}\Big)^{3/4}\Big(a^2+\frac{it}{m}\Big)^{-3/2}\mathtt{Exp}\Big[-\frac{\bm{x}^2}{2\big(a^2+\frac{it}{m}\big)}-imt\Big].
\end{equation}

Again, $e^{-i{\mathcal H}_{g}t}$ can also be further factorized by the BCH formula as 
\begin{align}\label{hg}
&e^{-i{\mathcal H}_{g}t}\sim e^{i\beta mgzt}e^{\frac{i\beta gt}{4m}\{{\bm p}^2,z\}}e^{\frac{i\beta gt}{4m}(\Sigma_2 p_1-\Sigma_1 p_2)}~~~~~\nonumber\\
&\quad=e^{i\beta mgzt}e^{\frac{i\beta gt}{4m}\{{\bm p}^2,z\}}e^{\frac{\beta gt}{4m}\Sigma_2 \partial_1}e^{\frac{-\beta gt}{4m}\Sigma_1 \partial_2}\nonumber\\
&\quad\sim e^{i\beta mgzt}( 1+\textstyle{\frac{i\beta gt}{4m}\{{\bm p}^2,z\}})e^{\frac{\beta gt}{4m}\Sigma_2 \partial_1}e^{\frac{-\beta gt}{4m}\Sigma_1 \partial_2}
\end{align}
Inserting Eq.(\ref{hg}) into Eq.(\ref{wf1}) by elementary calculation, we can obtain the approximation of the wavefunction $\psi_{\pm}(\bm{x},t)$ as follows
\begin{align}\label{wfg}
\psi_{\pm}(\bm{x},t)&\simeq\frac{1}{2}e^{i mgzt}( 1+\textstyle{\frac{igt}{4m}\{{\bm p}^2,z\}})\Big[\psi_{f\pm}(\bm{u},t)\nonumber\\
&\quad\quad+\psi_{f\pm}(\bm{v},t)+\psi_{f\mp}(\bm{v},t)-\psi_{f\mp}(\bm{u},t)\Big]
\end{align}
with 
\begin{align}
\bm{u}&=(x+\frac{\sigma gt}{4m},y+\frac{gt}{4m},z-\frac{1}{2}gt^2),\\
\bm{v}&=(x+\frac{\sigma gt}{4m},y-\frac{gt}{4m},z-\frac{1}{2}gt^2).
\end{align}
Equation (\ref{wfg}) confirms  the evolution towards an entangled state between the internal (spin) and external (position) degrees of freedom.

In fact, with the wavefunction for general time $t$ (\ref{wfg}), straightforward calculation can verify the expected position as 
\begin{equation}
\langle {\bm x}(t) \rangle_{\pm}=\big<\psi_{\pm}(\bm x,t)|\bm x|\psi_{\pm}(\bm x,t)\big>=(-\frac{\sigma gt}{4m},~0,~\frac{1}{2}g t^2)
\end{equation}
The probability distribution of position
\begin{equation}\label{rho}
|\psi_{\pm}(\bm{x},t)|^2\simeq\frac{1}{2}\big(|\phi_{f}(\bm{u},t)|^2+|\phi_{f}(\bm{v},t)|^2\big)
\end{equation}
In our case, since the mass dependence of $\bm{u}$ and $\bm{v}$ and the spin-position entanglement, the probability distribution is mass-dependent, and the functional mass-dependence of probability distribution is different from that of the free case in the absence of gravity. From this point, we can conclude that the notion of quantum WEP in Ref.~\cite{Anas18} is invalid, even in a static uniform gravitational field. 

In addition, from the view of the Fisher information framework, one can extract mass information of Dirac particle in free fall with spin polarization, even in a static uniform gravitational field, namely $F_x^g(m)\neq F_x^f(m)$. Thus it also imply the violation of the notion of quantum WEP in terms of Fisher information. In this sense, the gravitational spin Hall effect can be as a new probe to test the quantum version of WEP's notion.

\section{Discussion and Summary}\label{sec IV}

In conclusion, we have revealed the gravitational spin Hall effect holding simultaneously quantum and gravitational effects, even in a static uniform gravitational field. For the free-fall Dirac wave packets with opposite spin polarization, such remarkable effect suggests the transverse splitting in the direction perpendicular to spin orientation and gravity. In the F-W picture, we analyze the dynamical evolution of the free-fall Dirac wave packets and show that the gravitational spin Hall effect produces the entanglement between the internal and external degrees of freedom due to the gravitational spin-orbit coupling. Interestingly, the gravitational spin Hall effect manifest the violation of the quantum WEP's notion presented recently, even in the very simple case of a static uniform gravitational field.

To test the gravitational spin Hall effect will be of great interest and importance for the possible observation of the violation of WEP in the quantum realm. However, as showed previously, the shift is an order of magnitude smaller than the Compton wavelength of the Dirac particle. It does not seem feasible to detect this effect by the current detectors. If the future detectors realizing high spatial resolution and the free-fall particles traveling enough time, possible future observations of this effect could be as a new probe of WEP of quantum systems, so as to clarify the notion of quantum WEP.

\section*{Acknowledges}This work is supported by the Scientific Research Project of Hubei Polytechnic University (Project No. 20xjz02R).

\end{document}